\documentclass[doublecol]{epl2}

\title{Hamiltonian equation of motion and depinning phase transition in two-dimensional magnets}

\author{R. H. Dong$^{1}$, B. Zheng$^{1}$\footnote{corresponding author; email: zheng@zimp.zju.edu.cn} and N. J. Zhou$^{2}$}
\shortauthor{R. H. Dong, B. Zheng and N. J. Zhou}

\institute{
$^1$ Department of Physics, Zhejiang University, Hangzhou 310027,
P.R. China\\
$^2$ Department of Physics, Hangzhou Normal University, Hangzhou 310036, China}


\abstract{Based on the Hamiltonian equation of motion of the $\phi^4$ theory with quenched disorder,
we investigate the depinning
phase transition of the domain-wall motion in two-dimensional magnets.
With the short-time dynamic approach, we numerically determine the transition field, and the static and dynamic critical exponents.
The results show that the fundamental Hamiltonian equation of motion belongs to a universality class
very different from those effective equations of motion.}

\begin{document}

\maketitle

\section{Introduction}

In recent years the dynamics of elastic systems in disordered media has been a focus of theoretical and experimental studies \cite{he92,lem98,ono99,mou04,yam07,im09}.
Particularly, the magnetic domain-wall dynamics is an important topic in magnetic devices, nanomaterial, thin films, and semiconductor \cite{met07,shi07,dou08,kim09}.
At zero temperature, the domain-wall motion exhibits a depinning phase transition.
Due to the energy barriers created by the disorder, the domain wall is pinned and the velocity of the domain wall remains zero up to a transition field $H_{c}$ \cite{now98,due05,kol06a,bak08}.
At finite temperature, the phase transition is softened.

At present, most theoretical approaches to the domain-wall dynamics in ferromagnetic materials are typically based on the Edwards-Wilkinson equation with quenched disorder (QEW).
This equation is a phenomenological model and detailed microscopic structures and interactions of real materials are not concerned.
To understand the domain-wall motion at a microscopic level, one may build lattice models based on microscopic structures and interactions.
Very recently, with Monte Carlo simulations, the depinning transition and relaxation-to-creep transition have been carefully examined in
the two-dimensional (2D) random-field Ising model with a driving field (DRFIM) \cite{zho09,zho10,zho10a,don12}.
The results indicate that the QEW equation and DRFIM model are not in a same universality class, mainly due to the presence of
overhangs and islands in the DRFIM model. Both the QEW equation and Monte Carlo dynamics of the DRFIM model are effective theories,
although they may explain the experimental phenomena to some extent.
It is desirable to explore the fundamental theory of the domain-wall motion.

In general, it is believed that statistical mechanics is originated from the fundamental equations of motion
for many body systems, even though there has not been a general proof.
In statistical mechanics, ensemble theories and stochastic equations of motion are effective descriptions
of static and dynamic properties of the statistical systems, respectively.
In contrast, whether the fundamental equations of motion, e.g., Newton, Hamiltonian and Heisenberg equations etc,
could really produce the same results of statistical mechanics remains open
\cite{fer65,for92,esc94,ant95,cai98}. Numerical solutions of the Hamiltonian equations of the $\phi^4$ theory and $XY$ model, show that
both equilibrium and nonequilibrium properties of statistical systems could be described by the fundamental equation of motion,
at least for the order-disorder phase transition \cite{cai98,leo98,zhe99,zhe00a}.
The Hamiltonian equation of the $\phi^4$ theory falls into a same universality class of the Monte Carlo dynamics
of the Ising model.

The purpose of this paper is to examine the depinning phase transition in two-dimensional magnets,
with the Hamiltonian equations of motion of the 2D $\phi^4$ theory with quenched disorder,
and to compare the results with the effective dynamics such as the QEW equation and Monte Carlo dynamics.
The short-time dynamic approach is adopted to numerically extract the static and dynamic exponents.

\section{The model}

The Hamiltonian of the 2D $\phi^4$ theory with quenched disorder on a square lattice is
\begin{eqnarray}
\mathcal{H}  =   \sum_{i} [ \frac{1}{2}\pi_{i}^{2} + \frac{1}{2} \sum_{\mu} (\phi_{i+\mu} - \phi_{i})^{2} - \frac{1}{2} m^2\phi_{i}^2 \nonumber \\
  +  \frac{1}{4!} g\phi_{i}^4 - h_i\phi_i - H\phi_i ], \label{defH}
\end{eqnarray}
with  $\pi_{i}= \dot{ \phi_{i} }$.
The quenched random field $h_{i}$ is uniformly distributed within an interval $[-\Delta,\Delta]$, and $H$ is a homogeneous external driving field.
In this paper, we fix the parameters $m^2=6$, $g=1.8$ and $\Delta=6$. The Hamiltonian equation of motion is then written as
\begin{equation}
\ddot{\phi_i}= \sum_{\mu} (\phi_{i+\mu} + \phi_{i-\mu} -2\phi_i) + m^2\phi_{i} - \frac{g}{3!} \phi_{i}^3 + h_i + H. \label{equM}
\end{equation}
Since energy is conserved, solutions of the equation in the long-time regime are assumed to generate a microcanonical ensemble.
The temperature could be defined as the averaged kinetic energy in the long-time regime, but it may not be proper in the short-time regime.
Instead, the energy density could be the control parameter, since it is conserved and can be input from the initial state.
For $h_i=0$ and $H=0$, the ground state of the Hamiltonian in Eq.~(\ref{defH}) is given by $\pi_i=0$ and $\phi_i=c_0=\sqrt{6m^2/g}\approx 4.47$.

We consider a dynamic process with a semiordered initial state with a perfect domain wall in the $y$ direction.
As time evolves, the domain wall propagates and roughens, therefore it looks like an interface.
Simulations are performed on a square lattice, with a linear size $2L$ in the $x$ direction and $L$ in the $y$ direction.
Antiperiodic and periodic boundary conditions are adopted in the $x$ and $y$ directions, respectively.
Following Refs \cite{now98,rot01,zho09}, we rotate the square lattice such that the initial domain wall orients in the $(11)$ direction.
This avoids the pinning effect irrelevant for the quenched disorder.

To prepare a semiordered configuration, we first set the initial kinetic energy to zero, i.e., $\pi_{i}(0)=0$.
The configuration then consists of two fully ordered domains with the same magnitude $|\phi_i(0)| \equiv c_0$ and opposite signs.
The energy of the semiordered state differs from that of the ground state only by the domain wall.
In the limit of $L \to \infty$, this energy difference is infinitesimally small. In other words, the temperature is close to zero.

After the initial state is prepared, we simply discretize $\ddot{\phi}_i$ by $[\phi_i( t + \Delta t ) + \phi_i( t - \Delta t ) - 2\phi_i(t)]/(\Delta t)^2$,
and update the equation of motion up to $t=10^4$ with $\Delta t=0.02$.
Additional simulations with $\Delta t=0.05$ show that $\Delta t=0.02$ is sufficiently small.
At least $10000$ samples of disorder realization are used for average.
Most simulations are performed with $L=512$, and
simulations with different $L$ confirm that the finite-size effects are negligible.

To study the depinning transition of the domain interface, we first introduce a height function
\begin{equation}
h(y,t) = \frac{1}{L} \sum_{x=1}^L \phi_{xy}(t).
\label{defm}
\end{equation}
Here $\phi_{xy}(t)$ denotes the field at site $(x,y)$.
We may also map $\phi_{xy}$ to a Ising spin, i.e., $S_{xy}=Sgn(\phi_{xy})$, and then
define the height function as in the DRFIM model \cite{zho09,don12}.
Our simulations show that after a microscopic time scale $t_{mic}$, these two definitions yield the same results.

With the height function, the average velocity of the interface can be calculated
\begin{equation}
v(t) = \left \langle \frac{dh(y,t)}{dt}\right \rangle, \label{defv}
\end{equation}
where $<\cdots>$ includes the statistical average and average over
$y$. The roughness function is defined as
\begin{equation}
\omega^{2}(t) = \left \langle h(y,t)^2 \right \rangle -
\langle h(y,t) \rangle^2. \label{defw2}
\end{equation}

Let us assume that the depinning transition is of second order.
Based on the dynamic scaling arguments supported by the renormalization-group calculations, one expects that the order parameter $v$ obeys the scaling form \cite{jan89,zhe98,luo98,zhe99,zho09}
\begin{equation}
v(t, \tau, L) = b^{-\beta/\nu} v(b^{-z}t, b^{1/\nu}\tau,
b^{-1}L). \label{sclv1}
\end{equation}
Here $\beta$ and $\nu$ are the static exponents, $z$ is the dynamic exponent, and
$\tau=(H-H_c)/H_c$. The parameter $b$ is an arbitrary rescaling factor and $L$ is the lattice size.
In this dynamic system,
\begin{equation}
\xi(t) \sim t^{1/z}
\end{equation}
is assumed to be the nonequilibrium correlation length. Setting $b \sim \xi(t)$, the
dynamic scaling form is rewritten as
\begin{equation}
v(t, \tau, L) = t^{-\beta/\nu z} v(1, t^{1/\nu z}\tau, t^{-1/z}L). \label{sclv2}
\end{equation}
In the short-time regime, i.e., the regime with $\xi(t)
\ll L$, the finite-size effect is negligibly small,
\begin{equation}
v(t, \tau) = t^{-\beta / \nu z} v(t^{1/\nu z}\tau).
\label{sclv3}
\end{equation}
At the transition point $\tau = 0$, a power-law behavior
is expected,
\begin{equation}
v(t) = t^{-\beta / \nu z}. \label{sclv4}
\end{equation}
With Eq.~(\ref{sclv3}), the transition field $H_c$ may be located by
searching for the best power-law behavior of $v(t, \tau)$
\cite{zhe98,luo98,zho09}. The critical exponent $\beta /\nu z$ is then
estimated from Eq.~(\ref{sclv4}). To determine the critical exponent $1/\nu z$, we compute the logarithmic derivative of $v(t, \tau)$ \cite{zhe98,zho09}
\begin{equation}
\partial_{\tau}~ln~v(t,\tau)|_{\tau=0} \sim t^{1/\nu z}. \label{nu}
\end{equation}

As the domain wall propagates, it also roughen as an interface.
According to the definition of the roughness function $\omega^{2}(t)$ in Eq.~(\ref{defw2}), it includes the dynamic evolution of the bulk.
This may lead to a deviation from the scaling behavior of $\omega^{2}(t)$ in the short-time regime.
Therefore we redefine a pure roughness function $D\omega^{2}(t)$by eliminating the contribution from the bulk
\begin{equation}
D\omega^{2}(t)=\omega^{2}(t)-\omega_{b} ^{2}(t),
\end{equation}
where $\omega_{b} ^{2}(t)$ is actually the line susceptibility of the bulk in the $x$ direction.
$\omega_{b} ^{2}(t)$ can be computed from simulations starting from the fully ordered state.
In fact, $\omega_{b} ^{2}(t)$ quickly stabilizes to a constant.
Thus, for a sufficiently large lattice, the pure roughness function should scale as \cite{jos96,pan00,zho09,don12}
\begin{equation}
D\omega^{2}(t) \sim \xi(t)^{2\zeta}, \label{sclw2}
\end{equation}
where $\zeta$ is the global roughness exponent.

To independently estimate the exponent $z$, we introduce
\begin{equation}
F(t) = [M^{(2)}(t) - M(t)^2]/\omega^2(t). \label{defF}
\end{equation}
Here $M(t)$ is the global magnetization and $M^{(2)}(t)$ is
its second moment. In fact, $F(t)$ is nothing but the ratio of the
planar susceptibility and line susceptibility.
Since $\omega^2(t)$ describes the fluctuation in the $x$ direction and $M^{(2)}(t)-M(t)^2$ includes those in both $x$ and $y$ directions, the dynamic scaling behavior of $F(t)$ should be \cite{zhe98,zho09}
\begin{equation}
F(t) \sim \xi(t)/L \sim t^{1/z}/L \label{sclF}.
\end{equation}

A more informative quantity is the height correlation function
\begin{equation}
C(r, t) = \langle[h(y + r, t) - h(y, t)]^2\rangle.
\label{corF}
\end{equation}
Similar to the roughness function, one needs to eliminate the evolution of the bulk, $C_b(r,t)$.
The pure height correlation function $DC(r,t) = C(r,t) - C_b(r,t)$
should obey the standard scaling form
\begin{equation}
   DC(r,t) \sim  \left\{
   \begin{array}{lll}
    t^{2(\zeta - \zeta_{loc})/z}r^{2\zeta_{loc}}    &  &  \mbox{$r \ll \xi(t) $} \\
    t^{2\zeta/z}  &  &  \mbox{ $ \xi(t) \ll r$}
   \end{array}\right.
   \label{corS}
\end{equation}
with $\zeta_{loc}$ being the local roughness exponent.

\section{Numerical solutions}

\begin{figure}[t]
\includegraphics[scale=0.35]{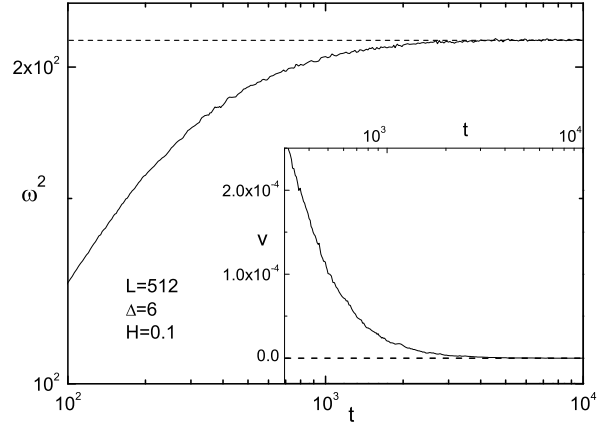}
\caption{The roughness function $\omega^2(t)$ is plotted at H=0.1 on a log-log scale.
In the inset, the velocity $v(t)$ is displayed on a semi-log scale. Dash lines represents constants.}\label{f1}
\end{figure}

The Hamiltonian equation of motion is deterministic in character, very different from the stochastic dynamics such as
the QEW equation and Monte Carlo simulation. We should first verify the existence of the pinning state.
In Fig.~\ref{f1}, the roughness function $\omega^2(t)$ and velocity $v(t)$ of the domain interface are plotted
as functions of $t$ for the disorder strength $\Delta=6$ with a weak driving field $H=0.1$.
Obviously, $\omega^2$ approaches a constant and $v$ drops down to zero within a certain period of time.
These results clearly indicate that the domain wall is pinned by the disorder for a weak driving field.
With a strong driving field $H$, the velocity will reach a constant, and the domain wall is depinned.
Inbetween these two states, there may exist a depinning phase transition.

In our simulations, three disorder strengths, $\Delta=2$, $4$ and $6$ have been examined.
However, for the weaker disorder, the depinning transition is
not typically of second order, it is likely of first order or in the crossover regime.
The power-law behavior in Eq.~(\ref{sclv4}) can not be identified.
This is similar to the DRFIM model with a weaker disorder \cite{zho09,zho10}.
Therefore, we concentrate our attention on $\Delta=6$, at which the depinning transition is of second order.

\begin{figure}[t]
\includegraphics[scale=0.35]{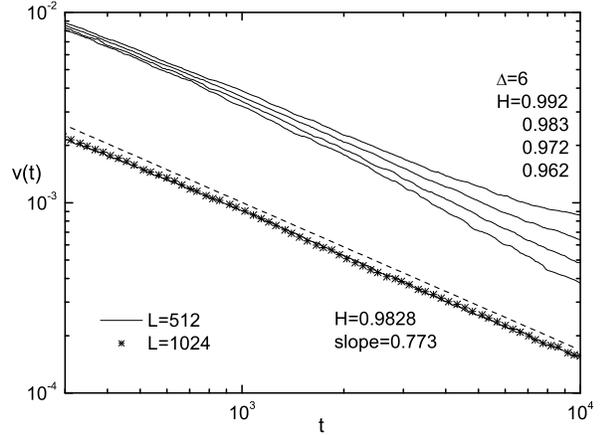}
\caption{The velocity v(t, $\tau$) is displayed for different driving fields $H$ on a log-log scale.
For clarity, the curves of different lattice sizes at $H_c=0.9828$ are shifted down. The dash line shows a power-law behavior at $H_c$.}\label{f2}
\end{figure}

In order to determine the transition field $H_c$, we perform simulations with different values of $H$
to search for the best power-law behavior of the velocity $v(t,\tau)$.
In Fig.~\ref{f2}, $v(t,\tau)$ is plotted as a function of $t$ for different driving field $H$ on a log-log scale.
A rather good power-law behavior is observed around $H=0.983$.
Careful analysis of the data locates the transition field $H_c=0.9828(3)$.
Both curves of $L=512$ and $1024$ nicely stabilize to a same power-law behavior,
suggesting that the finite-size effect is negligibly small.
The exponent $\beta / \nu z=0.773(3)$ is measured from the slope of the curve at $H_c$, according to Eq.~(\ref{sclv4}).

\begin{figure}[t]
\includegraphics[scale=0.35]{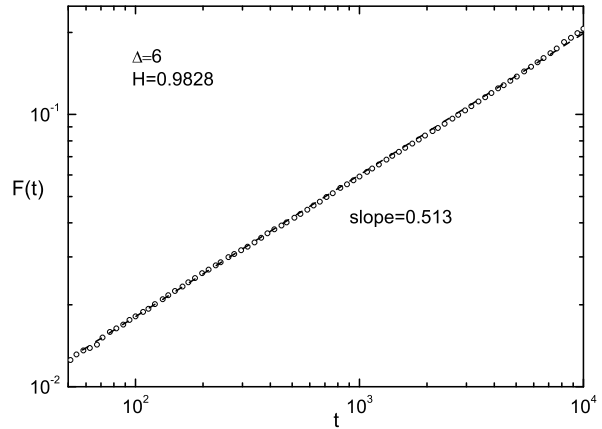}
\caption{$F(t)$ is plotted at $H_c=0.9828$ on a log-log scale. The dash line represents a power-law fit.}\label{f3}
\end{figure}

To extract the dynamic exponent $z$, $F(t)$ defined in Eq.~(\ref{defF}) is displayed at $H_c$ on a log-log scale in Fig.~\ref{f3}.
A power-law growth of $F(t)$ is observed from relatively early times.
Based on Eq.~(\ref{sclF}), a direct measurement of the slope of the curve gives $1/z=0.513(4)$.

\begin{figure}[t]
\includegraphics[scale=0.35]{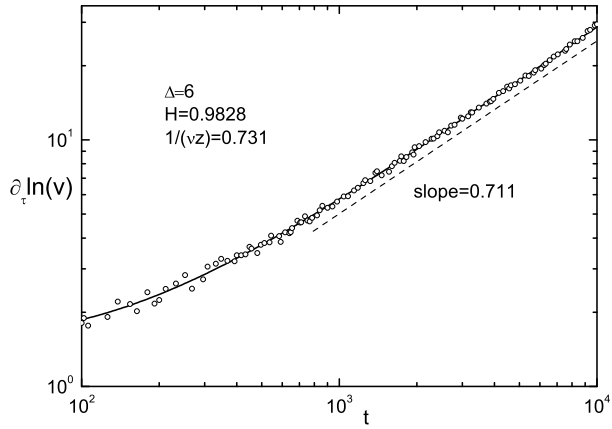}
\caption{The logarithmic derivative of $v(t, \tau)$ is displayed at $H_c=0.9828$ on a log-log scale. The dash line shows a power-law behavior,
and the solid line represents the fit with a power-law correction, $\partial_{\tau}~ln~v \sim t^{1/\nu z}(1+c/t)$.}\label{f4}
\end{figure}

To estimate the exponent $\nu$, we compute approximately the logarithmic derivative of the velocity in Eq.~(\ref{nu}).
In Fig.~\ref{f4}, $\partial_{\tau} \ln v(t,\tau)$ is plotted at $H_c$ on a log-log scale.
A power-law behavior is detected at later times,
and a direct measurement from the slope of the curve yields $1/\nu z=0.711(7)$.
Taking into account the corrections to scaling at early times,
we fit the curve to the form of $\partial_{\tau}~ln~v(t,\tau) \sim t^{1/\nu z}(1+c/t)$.
As shown in Fig.~\ref{f4}, the fitting to the curve is good, and it refines the measurement of the exponent to $1/\nu z=0.731(9)$.

From the measurements of $\beta/\nu z$, $1/\nu z$, and $1/z$, we compute the critical exponents
$\beta=1.06(1)$, $\nu=0.702(11)$ and $z=1.95(2)$.

\begin{figure}[t]
\includegraphics[scale=0.35]{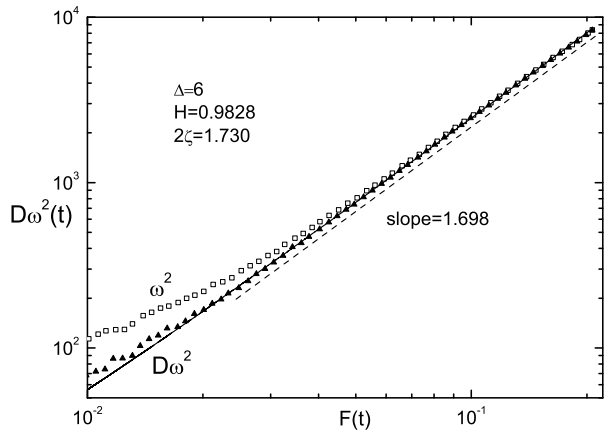}
\caption{The roughness function $\omega^2(t)$ and pure roughness function $D\omega^2(t)$
are plotted against $F(t)$ at $H_c=0.9828$ on a log-log scale. The dash line shows a power-law fit, while the solid line
includes a power-law correction to scaling.}\label{f5}
\end{figure}

From Eqs.~(\ref{sclw2}) and (\ref{sclF}), one expects the scaling form $D\omega^2(t) \sim F(t)^{2\zeta}$.
Our data analysis shows that determination of $\zeta$ from this scaling behavior is slightly better than from Eq.~(\ref{sclw2}).
In Fig.~\ref{f5}, therefore, the pure roughness function $D\omega^2(t)$ is plotted as a function of $F(t)$ at $H_c$ on a log-log scale.
A power-law behavior is observed after a transient regime.
From the slope of the curve one measures $\zeta=0.849(3)$.
To further refine the result, we consider a power-law correction in the form of $D\omega^2(t) \sim F(t)^{2\zeta}(1+c/F(t))$,
it leads to $\zeta=0.865(5)$. Compared with $D\omega^2(t)$, $\omega^2(t)$ obviously suffers from a stronger correction to scaling
at early times.

\begin{figure}[t]
\includegraphics[scale=0.35]{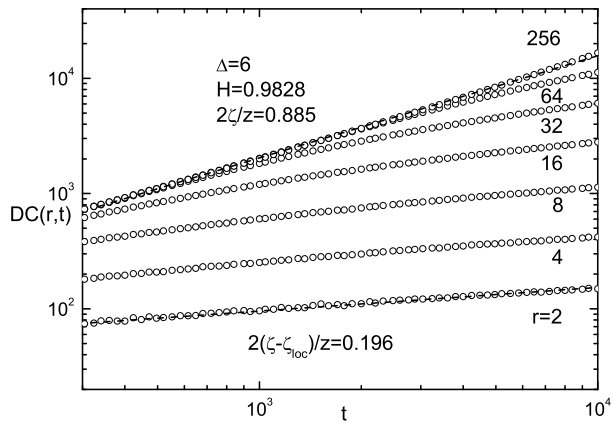}
\caption{The pure height correlation $DC(r,t)$ is displayed for different $r$ at $H_c=0.9828$ on a log-log scale.
Dash lines represent power-law fits.}\label{f6}
\end{figure}

In Fig.~\ref{f6}, the pure height correlation function $DC(r,t)$
is displayed for different $r$ at $H_c$. For a large $r \gg
\xi(t)$, e.g., $r = 256$, one extracts the exponent $2 \zeta / z =
0.885(6)$ by Eq.~(\ref{corS}), consistent with  $\zeta=0.865(5)$ from
Fig.~\ref{f5}. For a small $r \ll \xi(t)$, $DC(r,t)$ should be
independent of $t$ if $\zeta=\zeta_{loc}$,
according to Eq.~(\ref{corS}). In
Fig.~\ref{f6}, $DC(r,t)$ at $r=2$ clearly increases with $t$. A power-law
behavior is observed with an exponent $2 (\zeta - \zeta_{loc}) / z =
0.196(3)$.  One thus obtains $\zeta_{loc}=0.674(6)$. Additionally, one may also estimate $\zeta_{loc}$
from $DC(r,t) \sim r^{2\zeta_{loc}}$ at a large t, e.g., $t=10^4$.
The result $2 \zeta_{loc}=1.35(1)$ agrees with $\zeta_{loc}=0.674(6)$ from Fig.~\ref{f6}.

In Table \ref{t1}, all the critical exponents measured for the Hamiltonian equation of motion of the 2D $\phi^4$ theory with quenched disorder
are compared with those for the Monte Carlo dynamics of the 2D DRFIM model and the 2D QEW equation.
Obviously, the $\phi^4$ theory belongs to a universality class very different from that of
either the DRFIM model or the QEW equation.
The exponent $\beta$ of the $\phi^4$ theory is much larger.
By the definition in the steady state, the velocity $v(\tau) \sim \tau^\beta$. In the regime $H\geq H_c$, therefore, the external field easily drives
the domain wall propagating. 
The local roughness exponent $\zeta_{loc}$ of the $\phi^4$ theory is smaller than $1$,
similar to the DRFIM model, and different from the QEW equation. In the latter case, $\zeta_{loc} \approx 1$,
since the domain wall is simply assumed to be single-valued.
However, it is somewhat surprising that the global roughness exponent $\zeta$ of the $\phi^4$ theory is also smaller than $1$, different from
both the DRFIM model and QEW equation. In the DRFIM model, it is believed that overhangs and islands lead to $\zeta >1$.
For the QEW equation, it is a kind of inconsistence that $\zeta >1$ contradicts the single-valued domain wall.

\begin{figure}[t]
\includegraphics[scale=0.43]{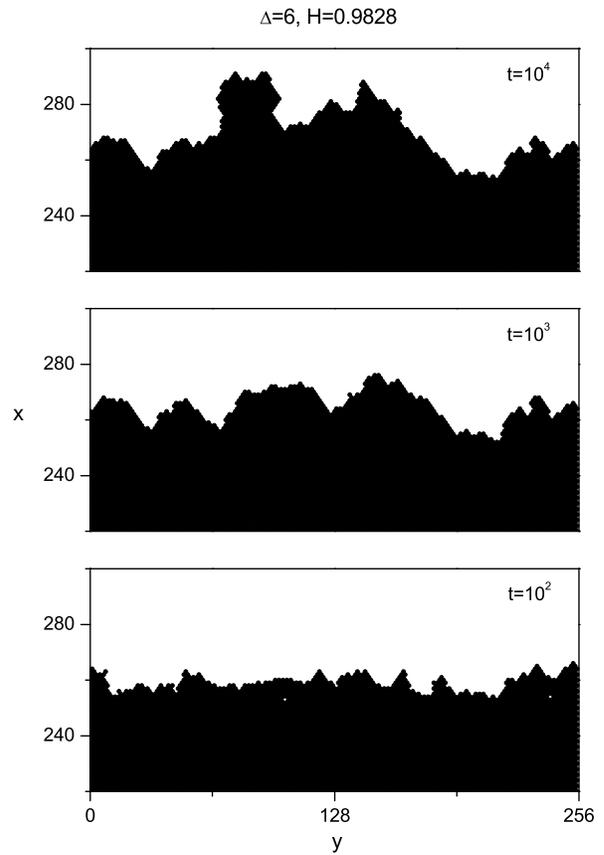}
\caption{The snapshot of the domain interface at $H_c=0.9828$. Black and white dots represent $Sgn(\phi_i)=1$ and $-1$ respectively.}\label{f7}
\end{figure}

To understand the roughening phenomena of the $\phi^4$ theory with quenched disorder,
the snapshot of the domain interface is displayed at $H_c$ in Fig. \ref{f7}.
As time evolves, the domain wall propagates and roughens. Overhangs are created, but not so prominent
as in the DRFIM model \cite{zho09}. Especially, islands can hardly be seen,
except for a number of small ones in the early times. Based on these facts, one may argue that the global roughness exponent $\zeta$
of the $\phi^4$ theory should be smaller than that of the DRFIM model. However, it remains somewhat puzzling
why it is smaller than $1$.

Experiments of the domain interface exactly at zero temperature do not exist. For $T>0$ and $0<H<H_c$,
it is reported that the local roughness exponent $\zeta_{loc}=0.7(1)$ and $0.69(7)$ in the
experiments with ultrathin Pt/Co/Pt films \cite{met07,lem98}, and
$\zeta_{loc}=0.78(1)$ with Co$_{28}$Pt$_{72}$ alloy films
\cite{jos98}. Our numerical value $\zeta_{loc}=0.674(6)$ is
compatible with these experimental results. We do not find direct experimental measurements of the global roughness exponent $\zeta$
of the domain interface, but other experiments such as those on kinetic surface roughening report results of both $\zeta > 1$ and $\zeta <1$
\cite{cor09}.

Finally we remind ourself that at the order-disorder phase transition,
the Hamiltonian equation of motion of the $\phi^4$ theory
falls into a same universality class of the Monte Carlo dynamics of the Ising model \cite{zhe99}.
Very probably this is because the order-disorder phase transition is an equilibrium phase transition,
and only relaxation dynamics is concerned.
However, the depinning phase transition of the domain-wall motion is a dynamic phase transition.
It is usually believed that the Monte Carlo dynamics may not quantitatively describe the dynamic transportation or propagation.
Therefore the Monte Carlo dynamics fails to produce the results of the Hamiltonian equation of motion
at the depinning phase transition.

\section{Conclusion}

We have numerically solved the Hamiltonian equation of motion of the 2D $\phi^4$ theory with quenched disorder, and carefully examined
the depinning phase transition of the domain-wall motion in two-dimensional magnets.
With the short-time dynamic approach, we determine the transition field $H_c$, and the static and dynamic critical exponents $\beta$, $\nu$, $z$, $\zeta$ and $\zeta_{loc}$.
The results show that the fundamental Hamiltonian equation of motion belongs to a universality class
very different from those effective equations of motion such as the QEW equation and the Monte Carlo dynamics of the DRFIM model.
In particular, the static exponent $\beta$ of the $\phi^4$ theory is much larger, and the global roughness exponent $\zeta < 1 $.
The local roughness exponent $\zeta_{loc} <1$ is compatible with experiments.

\acknowledgements This work was supported in part by NNSF of China
under Grant Nos. 10875102 and 11075137.


\begin{thebibliography}{10}
\expandafter\ifx\csname url\endcsname\relax\def\url#1{\texttt{#1}}\fi

\bibitem{he92}
\Name{{S.J. He, G.L.M.K.S. Kahanda, and P.Z. Wong}} \REVIEW{Phys. Rev. Lett.
}{69}{1992}{3731}.

\bibitem{lem98}
\Name{{S. Lemerle, J. Ferr\'{e}, C. Chappert, V. Mathet, T. Giamarchi, and P.
  Le Doussal}} \REVIEW{Phys. Rev. Lett.}{80}{1998}{849}.

\bibitem{ono99}
\Name{{T. Ono, H. Miyajima, K. Shigeto, K. Mibu, N. Hosoito, T. Shinjo}}
  \REVIEW{Science}{284}{1999}{468}.

\bibitem{mou04}
\Name{{S. Moulinet, A. Rosso, W. Krauth, and E. Rolley}} \REVIEW{Phys. Rev.
  E}{69}{2004}{035103(R)}.

\bibitem{yam07}
\Name{{M. Yamanouchi, J. Ieda, F. Matsukura, S.E. Barnes, S. Maekawa, and H.
  Ohno}} \REVIEW{Science}{317}{2007}{1726}.

\bibitem{im09}
\Name{{M.Y. Im, L. Bocklage, P. Fischer, and G. Meier}} \REVIEW{Phys. Rev.
  Lett.}{102}{2009}{147204}.

\bibitem{met07}
\Name{{P.J. Metaxas, J.P. Jamet, A. Mougin, M. Cormier, J. Ferr\'e, V. Baltz,
  B. Rodmacq, B. Dieny, and R. L. Stamps}} \REVIEW{Phys. Rev. Lett.
}{99}{2007}{217208}.

\bibitem{shi07}
\Name{{Y.H. Shin, I. Grinberg, I.W. Chen, and A.M. Rappe}} \REVIEW{Nature
}{449}{2007}{881}.

\bibitem{dou08}
\Name{{A. Dourlat, V. Jeudy, A. Lema\^{i}tre, and C. Gourdon}} \REVIEW{Phys.
  Rev. B}{78}{2008}{161303(R)}.

\bibitem{kim09}
\Name{{K.J. Kim, J.C. Lee, S.M. Ahn, K.S. Lee, C.W. Lee, Y.J. Cho, S. Seo, K.H.
  Shin, S.B. Choe, and H.W. Lee}} \REVIEW{Nature}{458}{2009}{740}.

\bibitem{now98}
\Name{{U. Nowak and K. D. Usadel}} \REVIEW{Europhys. Lett.}{44}{1998}{634}.

\bibitem{due05}
\Name{{O. Duemmer and W. Krauth}} \REVIEW{Phys. Rev. E}{71}{2005}{061601}.

\bibitem{kol06a}
\Name{{A.B. Kolton, A. Rosso, T. Giamarchi, and W. Krauth}} \REVIEW{Phys. Rev.
  Lett.}{97}{2006}{057001}.

\bibitem{bak08}
\Name{{B. Bak\'o, D. Weygand, M. Samaras, W. Hoffelner, and M. Zaiser}}
  \REVIEW{Phys. Rev. B}{78}{2008}{144104}.

\bibitem{zho09}
\Name{{N.J. Zhou, B. Zheng, and Y.Y. He}} \REVIEW{Phys. Rev. B
}{80}{2009}{134425}.

\bibitem{zho10}
\Name{{N.J. Zhou and B. Zheng}} \REVIEW{Phys. Rev. E }{82}{2010}{031139}.

\bibitem{zho10a}
\Name{{N.J. Zhou, B. Zheng and D.P. Landau}} \REVIEW{EPL}{92}{2010}{36001}.

\bibitem{don12}
\Name{{R.H. Dong, B. Zheng and N.J. Zhou}} \REVIEW{EPL}{98}{2012}{36002}.

\bibitem{fer65}
\Name{{E. Fermi, J. Pasta, and S. Ulam}} presented at \Book{{Collected Papers
  of Enrico Fermi}}, edited by \Name{{E. Segr\'e}} (Univ. Chicago, Chicago)
  1965.

\bibitem{for92}
\Name{{J. Ford}} \REVIEW{Phys. Rep.}{213}{1992}{271}.

\bibitem{esc94}
\Name{{D. Escande, H. Kantz, R. Livi, and S. Ruffo}} \REVIEW{J. Stat. Phys.
}{76}{1994}{605}.

\bibitem{ant95}
\Name{{M. Antoni and S. Ruffo}} \REVIEW{Phys. Rev. E}{52}{1995}{2361}.

\bibitem{cai98}
\Name{{L. Caiani, L. Casetti, and M. Pettini}} \REVIEW{J. Phys.
  A}{31}{1998}{3357}.

\bibitem{leo98}
\Name{{X. Leoncini and A.D. Verga}} \REVIEW{Phys. Rev. E}{57}{1998}{6377}.

\bibitem{zhe99}
\Name{{B. Zheng, M. Schulz, and S. Trimper}} \REVIEW{Phys. Rev. Lett.
}{82}{1999}{1891}.

\bibitem{zhe00a}
\Name{{B. Zheng}} \REVIEW{Phys. Rev. E}{61}{2000}{153}.

\bibitem{rot01}
\Name{{L. Roters, S. L\"ubeck, and K. D. Usadel}} \REVIEW{Phys. Rev.
  E}{63}{2001}{026113}.

\bibitem{jan89}
\Name{{H.K. Janssen, B. Schaub, and B. Schmittmann}} \REVIEW{Z. Phys.
  B}{73}{1989}{539}.

\bibitem{zhe98}
\Name{Zheng B.} \REVIEW{Int. J. Mod. Phys. B}{12}{1998}{1419} review article.

\bibitem{luo98}
\Name{{H.J. Luo, L. Sch\"ulke, and B. Zheng}} \REVIEW{Phys. Rev. Lett.
}{81}{1998}{180}.

\bibitem{jos96}
\Name{{M. Jost and K.D. Usadel}} \REVIEW{Phys. Rev. B}{54}{1996}{9314}.

\bibitem{pan00}
\Name{{N.N. Pang and W.J. Tzeng}} \REVIEW{Phys. Rev. E}{61}{2000}{3559}.

\bibitem{jos98}
\Name{{M. Jost, J. Heimel and T. Kleinefeld}} \REVIEW{Phys. Rev.
  B}{57}{1998}{5316}.

\bibitem{cor09}
\Name{{P. Co\'rdoba-Torres, T.J. Mesquita, I.N. Bastos, and R.P. Nogueira}}
  \REVIEW{Phys. Rev. Lett.}{055504}{2009}{102}.

\end{thebibliography}

\begin{table}
\caption{The critical exponents for the Hamiltonian equation of motion of the 2D $\phi^4$ theory are compared with
those for the Monte Carlo dynamics of the 2D DRFIM model and the 2D QEW equation.
The results for the 2D DRFIM model are taken from Ref. \cite{zho09}, and those for the 2D QEW equation
are summarized from Table I in Ref. \cite{zho09}.}
\label{t1}
 \centering \footnotesize
 \begin{tabular}{r|ccccc}
 \hline
 & $\beta$ & $\nu$ & $z$ & $\zeta$& $\zeta_{loc}$ \tabularnewline
\hline
$\phi^4$  & $1.06(1)$ & $0.70(1)$ & $1.95(2)$ & $0.865(5)$ & $0.674(6)$ \tabularnewline
 DRFIM  & $0.295(3)$ & $1.02(2)$ & $1.33(1)$ & $1.14(1)$ &$0.735(8)$ \tabularnewline
QEW& $0.33(2)$ & $1.33(4)$ &$1.50(3)$ & $1.25(1)$ & $0.98(6)$ \tabularnewline \hline
\end{tabular}

\end{table}

\end{document}